\documentclass[twocolumn,aps]{revtex4}

\begin{document}

 \title{Functional Bell inequalities can serve as a stronger entanglement witness}
 \author{Aditi Sen(De), Ujjwal Sen and Marek \.Zukowski}
 \affiliation{Instytut Fizyki Teoretycznej i Astrofizyki Uniwersytet
 Gda\'nski, PL-80-952 Gda\'nsk, Poland}

 \begin{abstract}

 We consider a Bell inequality for a \emph{continuous} range of settings of the
 apparatus at each site. This ``functional" Bell inequality gives a
 better range of violation for generalized GHZ states. Also
 a family of $N$-qubit bound entangled states violate this
inequality for $N>5$.

 \end{abstract}


 \maketitle

 \vspace{-0.5cm}

 A remarkable feature of entanglement
 is that it gives rise to correlations
 that cannot be explained by any local realistic theory.
 This is the statement of the Bell theorem \cite{Bell}. This theorem usually
 utilizes
 some inequalities that are satisfied by any local realistic theory
 but are violated by quantum correlations, between two or more systems. Modulo the well known
 loopholes,
 such violations have actually been experimentally demonstrated.
 Violation of such an inequality is a signature of entanglement. However it is
 not known
 whether the correlations in all entangled states are strong enough to
violate a Bell inequality.

 The usual formulation of the Bell theorem is for two apparatus settings at each
 site
 \cite{CHSH, MK, WWZB}.
 However there are several reasons for generalizing to more than two settings
 \cite{geq2, MZ1, ZK}.
 The simplest of them is that new Bell inequalities could reveal violation of local realism 
for cases when the standard inequalities fail.
 Or they could be more appropriate to some experimental situations
\cite{Franson}.

 In this paper, we consider a multipartite Bell inequality that involves a
 \emph{continuous}
 range of settings at each site
 (which we call the ``functional" Bell inequality).  We show that the functional
 Bell inequality is stronger in many cases than
 the standard Bell inequalities. This shows that this inequality may be a
 useful tool
 for
 classification of states with respect to  violation of local realism.

 All bipartite pure entangled states violate a Bell inequality \cite{Gisin1}.
 The multipartite situation is however more complicated. For example,
 using the
 Bell inequalities for correlation functions, which involve the usual choice between two observables for each of the local observers, the \(N\)-qubit generalised
 GHZ states
 \begin{equation}
 \label{GHZ}
  \left|\psi_N\right\rangle = sin\beta  \left|0\right\rangle^{\otimes N} +
 cos\beta
 \left|1\right\rangle^{\otimes N}
 \end{equation}
 (with \(0 \leq \beta \leq \pi/4\)) do not
 violate any such inequalities \emph{for \(N\) separated qubits
 without postselection}
 for \(sin2\beta \leq \frac{1}{\sqrt{2^{N-1}}}\) for odd \(N\) \cite{Gisin,
 MZ}.
 This is quite surprising, considering the fact that these states are a
 generalization
 of the GHZ state \cite{GHZ}
 \(\frac{1}{\sqrt{2}}\left(\left|0\right\rangle^{\otimes N} +
 \left|1\right\rangle^{\otimes N} \right)\)
 which strongly violates the standard Bell inequalities.
 For the functional Bell inequality that we consider in this paper,
 a violation is obtained for \(sin2\beta\geq 2\left(\frac{2}{\pi}\right)^N\)
which is better than the previous bounds for Bell violation for odd \(N\geq5\).

 One of the open questions in quantum information is whether bound entangled states
violate any Bell inequality. 
 Since the seminal works of dense coding \cite{BW} and teleportation
 \cite{BBCJPW}, the maximally
 entangled states have acquired special significance. However there exist
 entangled states which cannot be transformed into a maximally entangled state
 when the parties, sharing the entangled state, are separated. Such states has
 been
 called bound entangled states \cite{bound}. It is intriguing to consider
 whether such states can violate local realism. Indeed it has been conjectured
 that bound entangled states with positive partial transpose (PPT) \cite{PPT, Peres1}
 cannot violate local realism \cite{Peres}. Further work in this direction has
 been
 carried out
 in Refs. \cite{WW, MZ2}. In a recent paper, D{\"u}r \cite{Dur}, considers
 this question in the multipartite scenario. It is shown there that
 an \(N\)-qubit state
 \begin{equation}
 \label{Durstate}
 \rho_{N} = \frac{1}{N+1}\left(\left| GHZ\right\rangle\left\langle GHZ\right|
 + \frac{1}{2}\sum_{k=1}^{N}\left(P_k + \overline{P_k}\right)\right)
 \end{equation}
 violates a Mermin-Klyshko inequality \cite{MK}
 for \(N \geq 8\), despite being PPT in all \(1:N-1\) party cuts.
 Here
 \[\left|GHZ\right\rangle =
 \frac{1}{\sqrt{2}}\left(\left|0\right\rangle^{\otimes N} +
 e^{i\alpha_{N}}\left|1\right\rangle^{\otimes N}\right),
 \]
 with \(\alpha_N\) being a phase. And
 \(P_k = \left|\phi_k\right\rangle\left\langle\phi_k\right|\),
 \(\left|\phi_k\right\rangle = \left|0\right\rangle_1 \ldots
 \left|0\right\rangle_{k-1}\left|1\right\rangle_k\left|0\right\rangle_{k+1}
  \ldots \left|0\right\rangle_N\) with \(\overline{P_k}\) obtained from \(P_k\)
 by interchanging \(0\)s and \(1\)s in \(P_k\).
 The Bell violation
 in Ref. \cite{Dur} was exhibited for \(\alpha_N = \frac{\pi}{4(N-1)}\) \cite{notone}.
 Further work was recently done in Ref. \cite{Dagomir}, where violation of local
realism
was obtained for \(N\geq7\) for all values of the parameter \(\alpha_N\), by using Bell
 inequalities that involve three settings per observer.

 We show here that the functional Bell inequality is violated by the state
 \(\rho_N\)
for \(N\geq6\) irrespective of the value of the parameter \(\alpha_N\).

 To begin, let us discuss the functional Bell inequality \cite{MZ1}, which
 essentially follows from
 a simple geometric observation that
 in any real vector space, if for two vectors \(h\) and \(q\) one has
 \(\left\langle h \mid q\right\rangle < \parallel q \parallel^2\),
 then this
 immediately implies that \(h \ne q\). In simple words, if the
 scalar product of two vectors has a lower value than the length
of one of them, then the two vectors cannot be equal.

 Let \(\varrho_N\) be a state shared between \(N\) separated parties.
 Let \(O_n\) be an arbitrary observable at the \(n\)th location (\(n=1,\ldots,
 N\)).
 The quantum mechanical prediction \(E_{QM}\) for the
 correlation in the state \(\varrho_N\), when
 these observables are measured, is
 \begin{equation}
 \label{EQM}
 E_{QM}\left(\xi_1, \ldots, \xi_N\right) = Tr\left(O_1 \ldots O_N \varrho_N\right),
 \end{equation}
 where \(\xi_n\) is the aggregate of the local parameters at the \(n\)th site.
 Our object is to see whether this prediction can be reproduced
 in a local hidden variable theory. A local hidden variable correlation
 in the present scenario must be of the form
 \begin{equation}
 \label{EHV}
 E_{LHV}\left(\xi_1, \ldots, \xi_N\right) = \int d\lambda \rho (\lambda) \Pi_{n=1}
 ^{N}
 I_{n} ( \xi_{n}, \lambda),
 \end{equation}
 where \(\rho(\lambda)\) is the distribution of the local hidden variables and
 \( I_{n}(\xi_{n}, \lambda) \) is the predetermined measurement-result of the
 observable
 \(O_n(\xi_n)\) corresponding to the hidden variable \(\lambda\).

 Consider now the scalar product
 \begin{equation}
 \begin{array}{rcl}
 \displaystyle
 \left\langle E_{QM}\mid E_{LHV}\right\rangle = & \int & d\xi_1 \ldots
 d\xi_N \\
 & \times & E_{QM}\left(\xi_1, \ldots, \xi_N\right)
 E_{LHV}\left(\xi_1, \ldots, \xi_N\right)
 \end{array}
 \label{EQMEHV}
 \end{equation}
 and the norm
 \begin{equation}
 \label{normEQM}
 \parallel E_{QM} \parallel^2 =
 \int  d\xi_1 \ldots d\xi_N
 \left(E_{QM}\left(\xi_1, \ldots, \xi_N\right)\right)^{2}.
 \end{equation}
 If we can prove that a strict inequality holds, namely
 for all possible \(E_{LHV}\), one has 
\begin{equation}
\label{star}
\left\langle E_{QM}\mid E_{LHV} \right\rangle \leq B,
\end{equation}
with the number \(B < \parallel E_{QM} \parallel^2\), we would immediately have
 \(E_{QM} \ne E_{LHV}\), indicating that the correlations in
 the state \(\varrho_N\) are of a different character than in
 any local realistic theory. We then could say that the state \(\varrho_N\) violates
 the ``functional" Bell inequality (\ref{star}), as this 
Bell inequality is expressed in terms
 of a typical scalar product for square integrable functions. Note that the 
value of the product depends on a
 continuous
range of parameters (of the measuring apparatuses) at each site.

 Let us first consider the case of generalized GHZ states
 \(\left|\psi_N\right\rangle\) given by (\ref{GHZ}), and experiments in which
 each observer is allowed to measure the local observables
 \begin{equation}
 \label{observable}
  O_{n}(\phi_{n}) = \left| +, \phi_n \right\rangle \left\langle +, \phi_n \right|  -
 \left| -, \phi_n \right\rangle \left\langle -, \phi_n \right|,
 \end{equation}
 where
 \begin{equation}
 \label{eigvec}
 \left|\pm,\phi_n\right\rangle =
 \frac{1}{\sqrt{2}}\left(\left|0\right\rangle \pm
 e^{i\phi_{n}}\left|1\right\rangle\right).
 \end{equation}
 The aggregate \(\xi_n\) of local parameters at the \(n\)th site
is just the single parameter \(\phi_n\) here.

 With the notations introduced before, one can find that the correlation function
 \(E_{QM}\) for the generalised GHZ state is given by
\[
 \begin{array}{rcl}
 \displaystyle
  E_{QM}( \phi_{1}, \ldots, \phi_{N} ) & = & \left\langle \psi_N \right|
 O_{1} \ldots O_{N}
 \left|\psi_N \right\rangle \\
  & = & sin\left(2\beta\right) cos\left(\sum _{n=1} ^{N}\phi_{n}\right).
 \end{array}
 \]
 Suppose we want to reproduce this correlation function
 by local hidden variables. The corresponding correlation function must
 be of the form (\ref{EHV}).
 As the allowed values of the observable
 \(O_n(\phi_n)\) are \(\pm 1\), we must correspondingly have
\( I_{n}(\phi_{n}, \lambda) = \pm 1\).

 Defining the inner product between (the real-valued functions)
  \(f(\phi_1, \ldots, \phi_N)\)
 and \(g(\phi_1, \ldots, \phi_N)\) by
 \begin{equation}
 \label{innerp}\left\langle f \mid g \right\rangle =
 \Pi_{n=1}^{N}\left(\int_{0}^{2\pi}d\phi_n\right)f(\phi_1, \ldots, \phi_N)
 g(\phi_1, \ldots, \phi_N),
 \end{equation}
 we have (for the generalised GHZ state \(\left|\psi_N \right\rangle \))
 \[\parallel E_{QM}\parallel^{2} = \frac{(2\pi)^{N}}{2} sin^{2} 2 \beta, \]
 while
 \[
 \begin{array}{rcl}
 \displaystyle
 \left\langle E_{QM} \mid E_{LHV}\right\rangle
 & = & \Pi_{i=1}^{N}\left(\int_{0}^{2\pi} d\phi_{i}\right) \int d\lambda
 \rho(\lambda) \\
 & \times & {\Pi_{j=1}^{N}} I_{j}(\phi_{j}, \lambda) cos(\sum_{k=1}^{N}
 \phi_{k}) sin 2\beta.
 \end{array}
 \]
 It has been shown in Refs. \cite{MZ1, ZK} that the modulus of
 \[\Pi_{i=1}^{N}\left(\int_{0}^{2\pi} d\phi_{i}\right) \int d\lambda
 \rho(\lambda)
 {\Pi_{j=1}^{N}} I_{j}(\phi_{j}, \lambda) cos(\sum_{k=1}^{N} \phi_{k})\]
 is less than or equal to
 \(4^{N}\) (see eqns. (20-23) of Ref. \cite{MZ1} or eqns. (A9-A18) of
 \cite{ZK}).
 Consequently we have \(\left|\left\langle E_{QM} \mid
 E_{LHV}\right\rangle\right|
 \leq 4^{N}sin2\beta\) for the generalised GHZ state
\(\left|\psi_N\right\rangle\).

 Therefore
 \(\left|\left\langle E_{QM} \mid E_{LHV}\right\rangle\right|\)
 is strictly less than \(\parallel E_{QM}\parallel^{2}\) whenever
 \begin{equation}
 \label{relation}
  sin 2\beta > 2 \left( \frac{2}{\pi}\right)^{N}.
 \end{equation}
 So whenever
 eq. (\ref{relation})
 is satisfied, the generalised GHZ state \(\left|\psi_N\right\rangle\)
 violates the functional Bell inequality.
 Using the WWWZB Bell inequalities \cite{WWZB} 
for two settings per observer, the  range
 of violation is given by \(sin2\beta >
 \frac{1}{\sqrt{2^{N-1}}}\)
 for odd \(N\) \cite{Gisin, MZ}, and for even \(N\) violation is obtained for
 the whole
 range of
 \(\beta\) \cite{MZ}.
 Note that \(\frac{1}{\sqrt{2^{N-1}}} > 2 \left( \frac{2}{\pi}\right)^{N}\)
 \(\Leftrightarrow \left(\frac{\pi}{2^{3/2}}\right)^N > \sqrt{2}\) which holds
 for all
 \(N\geq4\).
 Therefore the functional Bell inequality
 shows a better range of violation for \emph{odd} \(N\geq5\).
And the difference between the two limits of \(\beta\) (for the standard Bell inequalities
and the functional Bell inequality) grows with \(N\).  
The region of violation covers the whole range of \(\beta\)
 as \(N\rightarrow \infty\) as in Ref. \cite{MZ}. 
However, note that the functional inequality is less
 restrictive in the case of $N$ even. Perhaps in this case a
different version of such an inequality must be used. We leave
this question open.

 Note that we only consider here
 violations of the inequality with all the \(N\) parties separated and without postselection.
 The state \(\left|\psi_N\right\rangle\) is just
 \(sin\beta\left|00\right\rangle
 + cos\beta\left|11\right\rangle\) in any bipartite cut with all the parties on
any side of the cut being together. Such a bipartite entangled
 state always  violate a Bell inequality \cite{Gisin1}.
 Also, allowing postselection would always result in a Bell violation as shown by
 Popescu and
Rohrlich \cite{PR}. Similar result can be shown using numerical approach \cite{Dagomir1}.

 We now go over
 to our second example of states given by (\ref{Durstate}). D{\" u}r \cite{Dur}
 obtained the following interesting result in the multi-qubit case. The state
 \(\rho_N\)
 of eq.(\ref{Durstate}) is
 PPT \cite{PPT} in all \(1:N-1\) party cuts. A bipartite state \(\rho_{AB}\)
 which is PPT can be either bound entangled or separable \cite{Peres1, bound}.
 If the state has negative partial transposition  \cite{PPT},
 it is always entangled \cite{Peres1}.
 The state \(\rho_N\) has a negative partial transpose for all \(2:N-2\) party cuts
 for
 \(N\geq4\). Consequently the state \(\rho_{N}\) (for \(N\geq4\))
 is a bound entangled state
 as long as all the parties are separated.
 Nevertheless, D{\" u}r \cite{Dur}
 showed that such states violate Mermin-Klyshko inequalities \cite{MK},
 with the
 allowed observables being between
 \(\sigma_x\) and \(\sigma_y\) at all the \(N\) locations (i.e. for the type given
 by our (\ref{observable})),
  for \(N\geq8\).
 Interestingly, Ac\'\i n \cite{Acin} showed that if a state violates
 the Mermin-Klyshko inequality, it would be possible to create a maximally
 entangled state in at least one bipartite cut. Further results were obtained in
 Ref. \cite{AcinScarani}.

In the case of states given by (\ref{Durstate}), 
again we allow each observer to make the measurements corresponding to the
 observables
 \( O _{n}( \phi _{n})\) defined in eq. (\ref{observable}).
 Then
 \[
 E_{QM} = Tr\left(O_1 \ldots O_N \rho_N\right) = \frac{1}{N+1}cos(\alpha_N -
 \sum_{i=1}^{N}
 \phi_i)
 \]
 Consequently (with the same inner product as in the previous case)
 \[
 \parallel E_{QM}\parallel^2 = \frac{1}{(N+1)^2}\frac{1}{2}(2\pi)^N
 \]
 and
\[ \left|\left\langle E_{QM}\mid E_{LHV}\right\rangle\right|
 \leq \frac{4^N}{N+1},
 \]
 which is again obtained from the results in Refs. \cite{MZ1, ZK}.
 And therefore
 \[
 \parallel E_{QM}\parallel^2 = \frac{1}{(N+1)^2}\frac{1}{2}(2\pi)^N
 > \frac{4^N}{N+1} \geq \left|\left\langle E_{QM}\mid
 E_{LHV}\right\rangle\right|
 \]
 i.e., \(E_{QM}\ne E_{LHV}\)
 for \(N\geq6\) irrespective of the value of the parameter
 \(\alpha_N\). Thus the state \(\rho_N\) violates
 the functional Bell inequality for \(N\geq6\) for all values of the parameter
 \(\alpha_N\).

 We have considered in this paper, a Bell inequality which seems to be
 effective when the standard Bell inequalities are not. Contrary
 to the standard ones, this Bell inequality
 is for a \emph{continuous} range of settings of the apparatus
 at each site. We have shown that this inequality shows Bell violation
 of the generalised GHZ state \(sin\beta\left|0\right\rangle^{\otimes N}
 + cos\beta\left|1\right\rangle^{\otimes N}\) (with \(N\) separated parties
 without postselection) for a larger range of the parameter \(\beta\) (for odd
 \(N\geq5\))
 than by
 any of the standard Bell inequalities.
 Further a \(6\)-qubit one-parameter family of bound entangled states
 is shown to violate this inequality.

 AS and US are supported by the University of Gda\'{n}sk,
 Grant No. BW/5400-5-0236-2. MZ is supported by the KBN grant
P 03B 088 20.


\begin{thebibliography}{9}
\bibitem{PPT}The partial
 transpose of a bipartite state \(\rho_{AB} = \rho_{i\mu j\nu}\), where
 \(i\) and \(j\) are the indices for party \(A\) and \(\mu\) and \(\nu\) are
 the indices for party \(B\), with respect to part \(A\) is \(\rho^{T_{A}}_{AB}
 = \rho_{j\mu i\nu}\) \cite{Peres1}. A state \(\rho_{AB}\) is
 said to have positive partial transpose (PPT) if \(\rho_{AB}^{T_A}\) is
 a positive operator. \(\rho_{AB}\) has negative partial transpose otherwise.
\bibitem{notone}Of course, one can show that the violation of the so-called
Mermin-Klyshko inequalities considered in \cite{Dur} occurs for all values
of \(\alpha_N\) for \(N \geq 8\).
\end{thebibliography}
 \end{document}